\documentclass[reprint,
superscriptaddress,
amsmath,amssymb,aps,prl
]{revtex4-2}
\usepackage{color,hyperref}
\hypersetup{colorlinks=true,linkcolor=blue,citecolor=red,filecolor=magenta,urlcolor=blue}
\usepackage{graphicx}
\usepackage{dcolumn}
\usepackage{bm}

\begin{document}

\preprint{APS/123-QED}

\title{Black Holes as a Collider of High Energy Particles}

\author{Bobur Turimov}\email{bturimov@astrin.uz}\affiliation{Ulugh Beg Astronomical Institute, Astronomy St. 33, Tashkent 100052, Uzbekistan}\affiliation{School of Engineering, Central Asian University, Tashkent 111221, Uzbekistan }\affiliation{New Uzbekistan University, Mustaqillik Avenue 54, Tashkent 100007, Uzbekistan} \author{Shuhrat Hayitov}\affiliation{Samarkand State University of Architecture and Construction, Lolazor St. 70, Samarkand 140147, Uzbekistan}

\date{\today}

\begin{abstract}
According to the Banados-Silk-West (BSW) process, rotating black holes can act as particle colliders capable of achieving arbitrarily high center-of-mass energy (CME), provided that a specific angular momentum of one of the particles is present. In this discussion, we demonstrate that both Kerr black holes and Schwarzschild black holes could serve as potential sources of high-energy particles in the polar region.
\end{abstract}

\maketitle

The origin of ultra-high-energy cosmic rays (UHECRs) is still unclear and their observational evidence of energy $10^{20} \rm eV$ posses some interesting and challenging questions in the scientific community. It is widely believed that they originate from extragalactic sources \cite{Kotera2011ARAA}. The observed spectrum of such UHECRs exhibits several distinctive features \cite{Hillas1984ARAA,Rodrigues2021PRL}. At approximately $4\times 10^{18} \rm eV$, there is a hardening in the spectrum known as the "ankle." This phenomenon can be attributed to a transition from Galactic to extragalactic cosmic rays (CRs) in models with either mixed composition or iron dominance \cite{Allard2007APh}. Alternatively, in proton-dominated models, the hardening could arise from losses due to pair production during propagation \cite{Berezinsky2006PRD}. 

From a theoretical perspective, there are several models that explain the mechanisms behind high-energy particles. For example, the Blandford-Znajec mechanism \cite{Blandford1977MNRAS} elucidates the release of rotational energy from a rotating black hole. Another model, known as the Penrose process \cite{Penrose1971NPhS}, describes how energy can be extracted from a rotating black hole, with a maximum energy efficiency of approximately $21\%$. In the magnetic Penrose process, this efficiency exceeds $100\%$ \cite{Dadhich2018MNRAS}. It has also been investigated the annihilation of dark matter particles in the gravitational field of black holes \cite{Baushev2009IJMPD} and shown that the CME of a pair of colliding identical particles in this scenario is given by $E_{c.m.}=2\sqrt{5}m$. Additionally, the BSW process \cite{BSW2009PRL} provides insight into collisions between pairs of particles falling into rotating black holes. Interestingly, for specific values of the angular momentum (i.e., $l_1=2$ or $l_2=2$) of one of the particles, this process can generate arbitrarily large CME near the horizon. However, this model showed that the static black hole can not release high energy.

\begin{figure}
\includegraphics[width=\hsize]{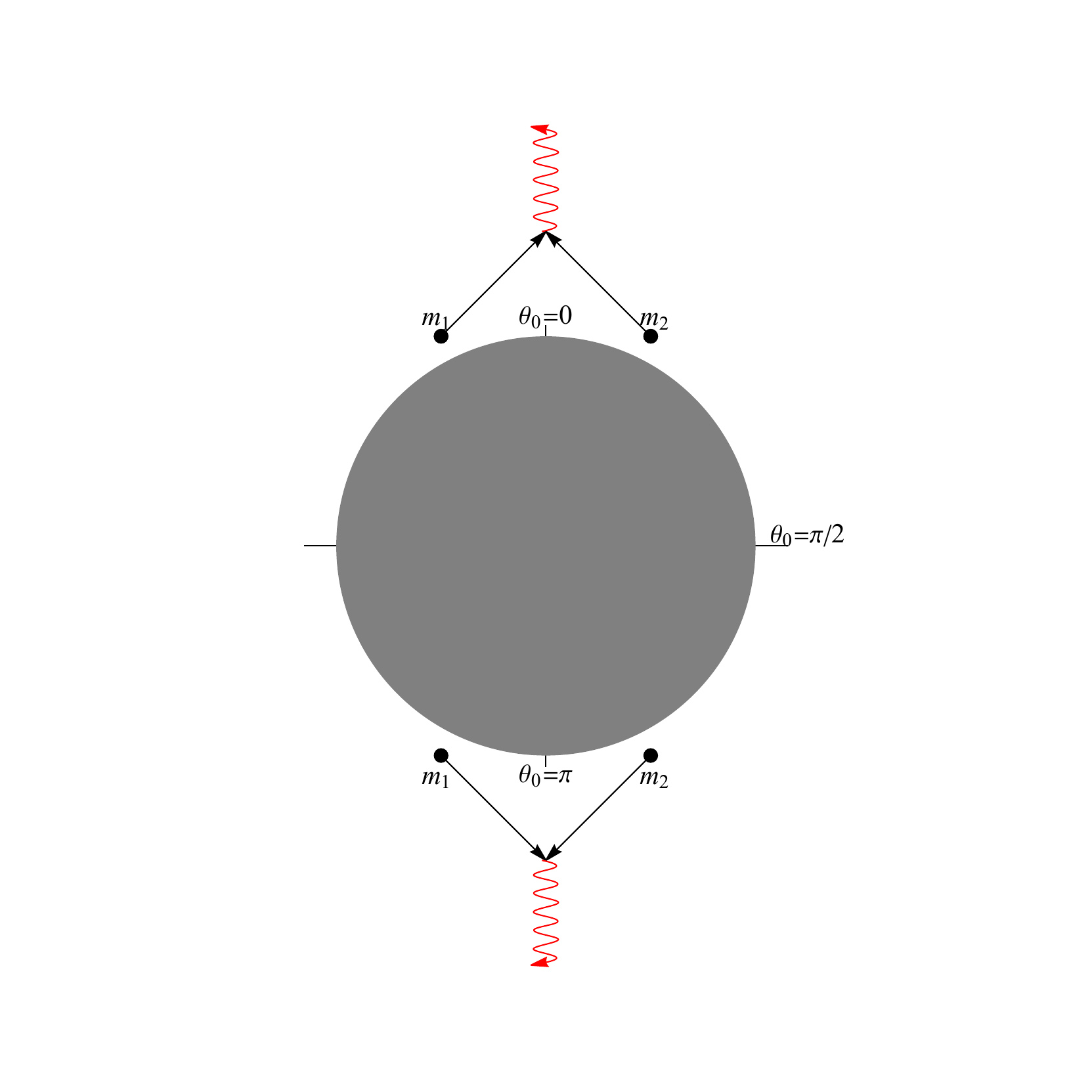}\caption{The schematic picture of colliding pair of particles in the polar region of the black hole.\label{schem}}
\end{figure}

In this Letter, we have conducted an investigation into a model aimed at elucidating the release of high-energy relativistic particles from the polar region of black holes. Our model builds upon the BSW process by incorporating the angular motion of test particles around both Kerr and Schwarzschild black holes. Remarkably, we have found that the CME (Collisional Mass-Energy) of colliding particle pairs diverges towards infinity in the polar region of both types of black holes. This intriguing phenomenon offers a potential explanation for the mechanisms behind high-energy sources observed in astrophysics, including Quasars, Blazars, and other similar sources. For a visual representation of this process, refer to the schematic illustration depicted in Figure \ref{schem}.
\begin{figure}
\includegraphics[width=\hsize]{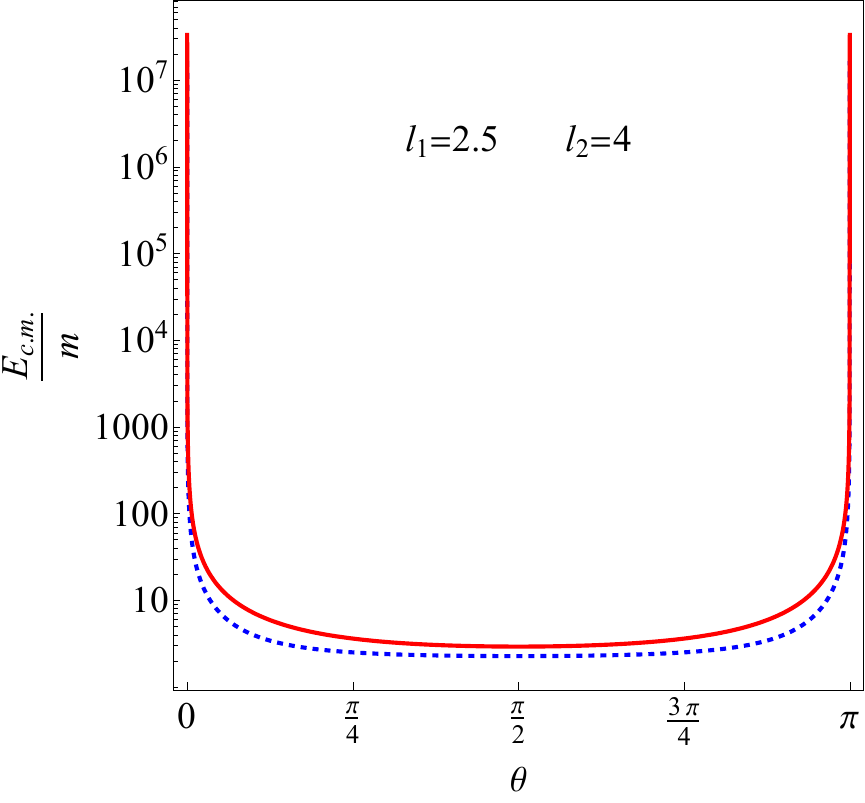}\caption{The angular dependence of the CME colliding particles near the event horizon of an extremal Kerr black hole (solid red line) and a Schwarzschild black hole (dashed blue line) for a specific value of angular momentum of particles  .\label{CME}}
\end{figure}

\begin{figure}
\includegraphics[width=\hsize]{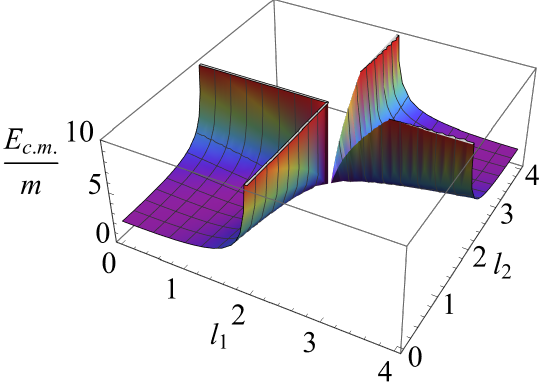}\caption{The existence of CME of colliding pair of identical particles in the equatorial plane ($\theta_0=\pi/2$).\label{Region}}
\end{figure}

Ignoring the effects arising from limitations on the maximal spin of the black hole, back-reaction effects, and sensitivity to the initial conditions of the collisions, as mentioned in \cite{Berti2009PRL}, we focus on the collision of a pair of massive particles near Kerr black holes. However, it is important to note that these effects may be considered in future studies. The main objective of this Letter is to provide a qualitative analysis of particle collisions near a black hole. According to Ref. \cite{BSW2009PRL}, the center-of-mass energy (CME) for the colliding pair of particles in curved spacetime is given by $E_{\rm c.m.}^2 = -g^{\mu\nu}(p_{1\mu}+p_{2\mu})(p_{1\nu}+p_{2\nu})$. Here, $p_{i\mu} = m_i{\dot x}_{i\mu}$ represents the four-momentum of the particles, satisfying $p_{i\mu}p^{i\mu} = -m_i^2$, and ${\dot x}_i^\mu$ denotes the four-velocity of the colliding particles with mass $m_i$ (where $i=1,2$). Finally, the CME of the colliding pair of particles is given by:
\begin{align}\label{E0}
\frac{E_{\rm c.m.}^2}{2m_1m_2}=1+\frac{(m_1-m_2)^2}{2m_1m_2}-g_{\mu\nu}{\dot x}_{1}^{\mu}{\dot x}_{2}^{\nu}\ .    
\end{align}
Here we focus this equation in the Kerr spacetime. The equations of motion in Kerr spacetime given as \cite{Carter1968PR}
\begin{align}\label{eq1}
&\Sigma{\dot r}=\sqrt{\left[(r^2+a^2){\cal E}-al\right]^2-\Delta\left({\cal K}+r^2\right)}\ ,
\\\label{eq2} 
&\Sigma{\dot\theta}=\sqrt{{\cal K}-\frac{(l-a{\cal E}\sin^2\theta)^2}{\sin^2\theta}-a^2\cos^2\theta}\ ,    
\\\label{eq3}
&\Sigma{\dot\phi}=\frac{a}{\Delta}[(r^2+a^2){\cal E}-al]+\frac{l-a{\cal E}\sin^2\theta}{\sin^2\theta}\ ,
\\
&\Sigma{\dot t}=\frac{r^2+a^2}{\Delta}[(r^2+a^2){\cal E}-al]+a(l-a{\cal E}\sin^2\theta)\ , \label{eq4}
\end{align}
and associated with three constants of motion, specific energy ${\cal E}$, specific angular momentum $l$, and Carter constant ${\cal K}$, where $\Delta=r^2-2r+a^2$, $\Sigma=r^2+a^2\cos^2\theta$. Using equation of motion \eqref{eq1}-\eqref{eq4}, the CME in the background of the Kerr spacetime takes a form:
\begin{widetext}
\begin{align}\label{Ecm2}\nonumber
\frac{E_{\rm c.m.}^2}{2m_1m_2}=1&+\frac{(m_1-m_2)^2}{2m_1m_2}+\frac{R_1(r)R_2(r)-\sqrt{R_1^2(r)-\Delta(r^2+{\cal K}_1)}\sqrt{R_2^2(r)-\Delta(r^2+{\cal K}_2)}}{\Sigma\Delta}\\&-\frac{T_1(\theta)T_2(\theta)+\sqrt{{\cal K}_1-T_1^2(\theta)-a^2\cos^2\theta}\sqrt{{\cal K}_2-T_2^2(\theta)-a^2\cos^2\theta}}{\Sigma}\ ,    
\end{align}
\end{widetext}
where $R_i(r)=(r^2+a^2){\cal E}_i-al_i$ and $T_i(\theta)=l_i\csc\theta-a{\cal E}_i\sin\theta$. As evident from the analysis of Eq.\eqref{Ecm2}, a singularity arises at the horizon of the black hole (i.e., $r_+=1+\sqrt{1-a^2}$ or $\Delta=0$). 
However, this issue can be resolved by introducing a modification. By multiplying the expression $R_1(r)R_2(r)+\sqrt{R_1^2(r)-\Delta(r^2+{\cal K}_1)}\sqrt{R_2^2(r)-\Delta(r^2+{\cal K}_2)}$ to both the numerator and denominator of the third term in equation \eqref{Ecm2} and subsequently applying the condition $\Delta=0$, the singularity can be mitigated. Nevertheless, there exist two additional singularities at $\theta=0$ and $\theta=\pi$ due to the presence of the trigonometric function $\csc\theta$ in the last term of equation \eqref{Ecm2}, which cannot be avoided. Consequently, it can be concluded that the center-of-mass energy (CME) of colliding particles tends to infinity in the polar region of the black hole, not only near the horizon but also at other points within these regions. It is important to note that equations of motion \eqref{eq1}-\eqref{eq4} are dependent on three constants, and for simplicity, we can set ${\cal E}_i=1$ for particles approaching the black hole from infinity \cite{BSW2009PRL}. Furthermore, at $\theta=0, \pi$, equation \eqref{eq2} exhibits divergence, indicating that the Carter constant approaches infinity. Hence, for a specific angle $\theta_0$ at which particles collide, including $0$ and $\pi$, the following condition can be obtained: ${\cal K}_i=(l_i-a\sin^2\theta_0)^2\csc^2\theta_0+a^2\cos^2\theta_0$. Substituting this condition into \eqref{Ecm2}, we can derive 
\begin{align}\label{Ecm}\nonumber
\frac{E_{\rm c.m.}^2}{m_1m_2}&=\frac{(m_1-m_2)^2}{m_1m_2}+\frac{[R_1(r_+)+R_2(r_+)]^2}{R_1(r_+)R_2(r_+)}\\&+\frac{[R_1(r_+)T_2(\theta_0)-R_2(r_+)T_1(\theta_0)]^2}{\Sigma(r_+,\theta_0)R_1(r_+)R_2(r_+)}\ ,    
\end{align}
which depends on black hole's horizon, the angle that particle collide, as well as their angular momentum. 
Near the horizon of the extrmal Kerr black hole with $r_+=1$ and $a=1$, equation \eqref{Ecm} yields
\begin{align}
\frac{E_{\rm c.m.}^2}{m_1m_2}=\frac{(m_1+m_2)^2}{m_1m_2}+\frac{2(l_2-l_1)^2}{(l_1-2)(l_2-2)\sin^2\theta_0}\ ,    
\end{align}
while near the horizon of the Schwarzschild black hole it reduces to
\begin{align}
\frac{E_{\rm c.m.}^2(r\to 2)}{m_1m_2}=\frac{(m_1+m_2)^2}{m_1m_2}+\frac{({l}_2-{l}_1)^2}{2\sin^2\theta_0}\ .    
\end{align}
Notice that by setting $\theta_0=\pi/2$ and $m_1=m_2=m$, one can reproduce results for the CME that is reported in \cite{BSW2009PRL}. Now one can see that the CME depends on not only angular momentum of particles but also the angle $\theta_0$ that particles collide. The CME of identical particles with identical angular momentum is determined as $E_{\rm c.m.}=2m$. Interestingly, in the polar region of the black hole i.e. $\theta_0=0, \pi$, the CME tends to infinity when angular momentum of particles are different from each others. Figure \ref{CME} represents angular dependence of the CME of pair colliding particles, while Figure \ref{Region} draws possible way of obtaining the real CME for particles in the $(l_1,l_2)$ plane at $\theta_0=\pi/2$. It shows CME will be always real when $l_i<2$ or $l_i>2$. It is observed that the CME remains real when $l_i<2$ or $l_i>2$. Additionally, it is worth considering the impact of factors such as the maximum spin limit of the black hole, back-reaction effects, and the sensitivity to initial collision conditions. Nevertheless, the main result for the CME in Eq. \eqref{Ecm2} contains a singularity at $\theta_0=0, \pi$.

\begin{acknowledgments}
This research is supported by Grants F-FA-2021-510 of the Uzbekistan Ministry for Innovative Development.
\end{acknowledgments}

\nocite{*}
\bibliography{Ref}
\end{document}